\documentclass{elsart}
\usepackage{natbib}
\usepackage{amssymb}
\usepackage{graphicx}

\newcommand{\be}  { \begin{equation} }
\newcommand{\ee}  { \end{equation}   }
\newcommand{\bea} { \begin{eqnarray} }
\newcommand{\eea} { \end{eqnarray}   }

\begin{document}
\begin{frontmatter}
\title{To Constrain the Parameter Space of the ICS Model}
\author{X. H. Cui$^1$}, 
\ead{xhcui@bac.pku.edu.cn}
\author{K. J. Lee$^1$},
\author{G. J. Qiao$^1$},
\author{Y. L. Yue$^1$\corauthref{cor1}}\corauth[cor1]{Corresponding author.},
\ead{yueyl@bac.pku.edu.cn}
\author{R. X. Xu$^1$}
\author{H. G. Wang$^2$}
\address{$^1$ Astronomy Department, School of Physics, Peking University,
         Beijing 100871, China}
\address{$^2$ Center for astrophysics, Guangzhou
University, Guangzhou 510400, China, cosmic008@263.net}

\begin{abstract}
The pulse profile of pulsar gives geometric information about
pulsar's radiation model. After investigating the pulse profiles of
PSR B1642-03 and PSR B0950+08, we calculate the ratios of beam width
and the emission height between different frequencies. We find that
the ratios are almost constants as inclination angle $\alpha$
changes from $0^{\circ}$ to $90^{\circ}$. The ratios can be used to
test pulsar's radiation model. In particular it can well constrain
the parameter space of the inverse Compton scattering (ICS) model.
These constrained parameters indicate some physical implication for
the ICS model.
\end{abstract}


\end{frontmatter}
%\section{Introuduction}

\section{Introduction}

Radhakrishnan and Cooke (1969) proposed the rotation vector model
(RVM) to explain the `S'-shape polarization position angle curve of
radio pulsar. Basing on his model and the relation between beam
radius $\rho$ and the period P, $\rho\sim P^{-1/3}$, Lyne and
Manchester (1988, hereafter LM88) calculated the value of
inclination angle $\alpha$ for over 200 pulsars. Applying another
$\rho-P$ relation: $\rho\sim P^{-1/2}$ and RVM, Rankin (1993) gave
the values of $\alpha$ for 150 pulsars. However, the exact value of
$\alpha$ can not be well determined for the pulsars with narrow
pulse profiles, e.g. PSR 0525+21 (Gould \& Lyne 1998). In this way,
to determine the physical or geometrical parameters, which in
sensitive to $\alpha$, becomes a remarkable problem. However we find
that the ratio between the beam widths at given two frequencies are
almost a constant respective to the changing of $\alpha$, and so do
the ratio between the height of radiation location at two
frequencies. Thus we can use this approximately unaltered ratio to
test the radiation models. In this paper, we use the results to
constraint the parameter space of inverse Compton scattering (ICS)
model (Qiao, 1988) and give the physical implication.

\section{The Observation and the Calculation}

Kramer (1994) studied the pulse profiles of PSR B1642-03 and PSR
B0950+08 and gave the pulse widths at three different frequencies.
For PSR B0950+08, the pulse width increases firstly and then
decreases with the increasing of frequency. But for PSR B1642-03, it
only increases while the increasing of frequency. LM88 gave the
maximum slope $\kappa$ of the position-angle curve.

The $\kappa$ is given below in RVM.
\begin{equation}
(d\psi/d\phi)_\mathrm{m} =\sin\alpha/\sin\beta\equiv\kappa,
\label{Max}
\end{equation}
where $\psi$ and $\phi$ are the polarization position angle and the
phase, $\beta$ is the impact angle, that is, the minimum angle
between line of sight and the magnetic axis. LM88 shows that
$\kappa=50$ and $1.4$ for PSR B1642-03 and PSR B0950+08
respectively, which are used to calculate the $\beta$ in this paper.
Next we will show that the radio between the beam widths and the
heights of radiation location at different frequencies is nearly a
invariant respective to $\alpha$.

The beam width $\theta_{\mu}$ can be obtained from the observed pulse width $\Delta\phi$ by
\begin{equation}
\sin^2(\theta_{\mu}/2)=\sin^2(\Delta\phi/2)\sin\alpha\sin(\alpha+\beta)+\sin^2(\beta/2).
\label{rho}
\end{equation}

Now we calculate the height of radiation location. If the emission
in different frequency comes from different magnetic field, the
phase $\varphi$ of a field line in the magnetic frame system as
shown in Fig.1 can be obtained from
$\cos(\alpha+\beta)=\cos\alpha\cos\theta_{\mu}-\sin\alpha\sin\theta_{\mu}\cos\varphi$.

\begin{figure}
\begin{center}
\includegraphics*[totalheight=2.5in]{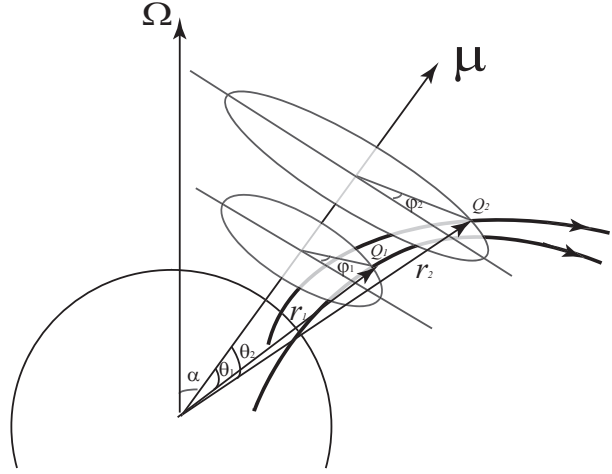}
\end{center}
\caption{Sketch of the magnetic field line, emission beams, and the
emission beams (two ellipses) in two frequencies: $f_1$ and $f_2$
$Q_1$ and $Q_2$ are two emission points. $r_1$, $\theta_1$,
$\varphi_1$ and $r_2$, $\theta_2$, $\varphi_2$ are the emission
heights, polar angles, and phases of $Q_1$ and $Q_2$, respectively.
\label{fig:1}}
\end{figure}
The maximum radius of a given magnetic field lines, i.e. the
furthest distance from the magnetic field line to the neutron star,
is
\begin{equation}
Re=\frac{R_{\mathrm{LC}}}{\sin^2\theta_\mathrm{M}\sqrt{1-(\cos\alpha\cos\theta_\mathrm{M}-\sin\alpha\sin\theta_\mathrm{M}\cos\varphi)^2}},
\label{Re}
\end{equation}
where $R_{\mathrm{LC}}$ is the radius of light cylinder and
$\theta_\mathrm{M}$ is the angle between the radius and the rotation
axis when the projection of radius in the vertical plane to the
rotation axis is equal to $R_{\mathrm{LC}}$. Then the
emission height $r$ are given by the equation of magnetic field:
$r=Re\sin^2\theta$, $\theta$ is the polar angle at point Q as shown in Fig.1.

From the three observed pulse widths at three frequencies from
Kramer (1994) and used the $\kappa$ from LM88, the three relations
of $\theta_{\mu 2}/\theta_{\mu 1}$ and $r_2/r_1$ as the function of
$\alpha$ for PSR B1642-03 and PSR B0950+08 are calculated by above
method and are shown in Fig.2 and Fig.3.

\begin{figure}
\begin{center}
\includegraphics*[width=.8\textwidth]{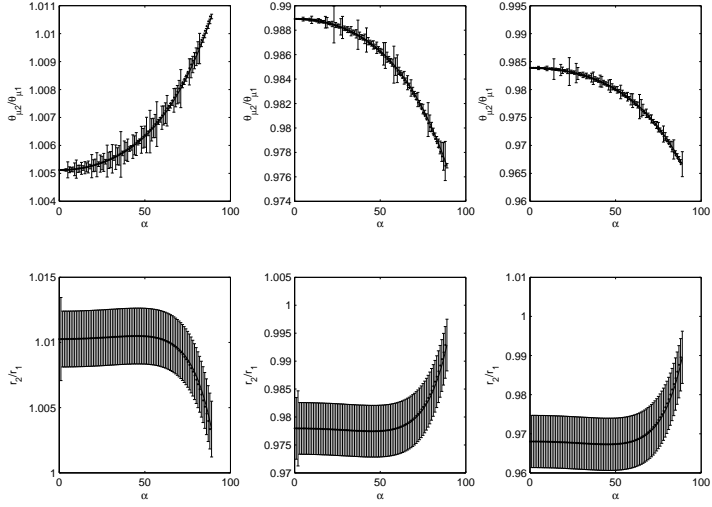}
\end{center}
\caption{$\theta_{\mu2}/\theta_{\mu1}$ and $r_2/r_1$ as the function
of $\alpha$ for PSR B0950+08. The left, middle, and right panels are
the ratios of 4.75GHz to 1.43GHz,  10.55GHz to 1.43GHz, and 10.55GHz
to 4.75GHz. The error bars indicate the error passing from the
errors of $\Delta\phi$ and $\kappa$. \label{fig:2}}
\end{figure}
\begin{figure}
\begin{center}
\includegraphics*[width=.8\textwidth]{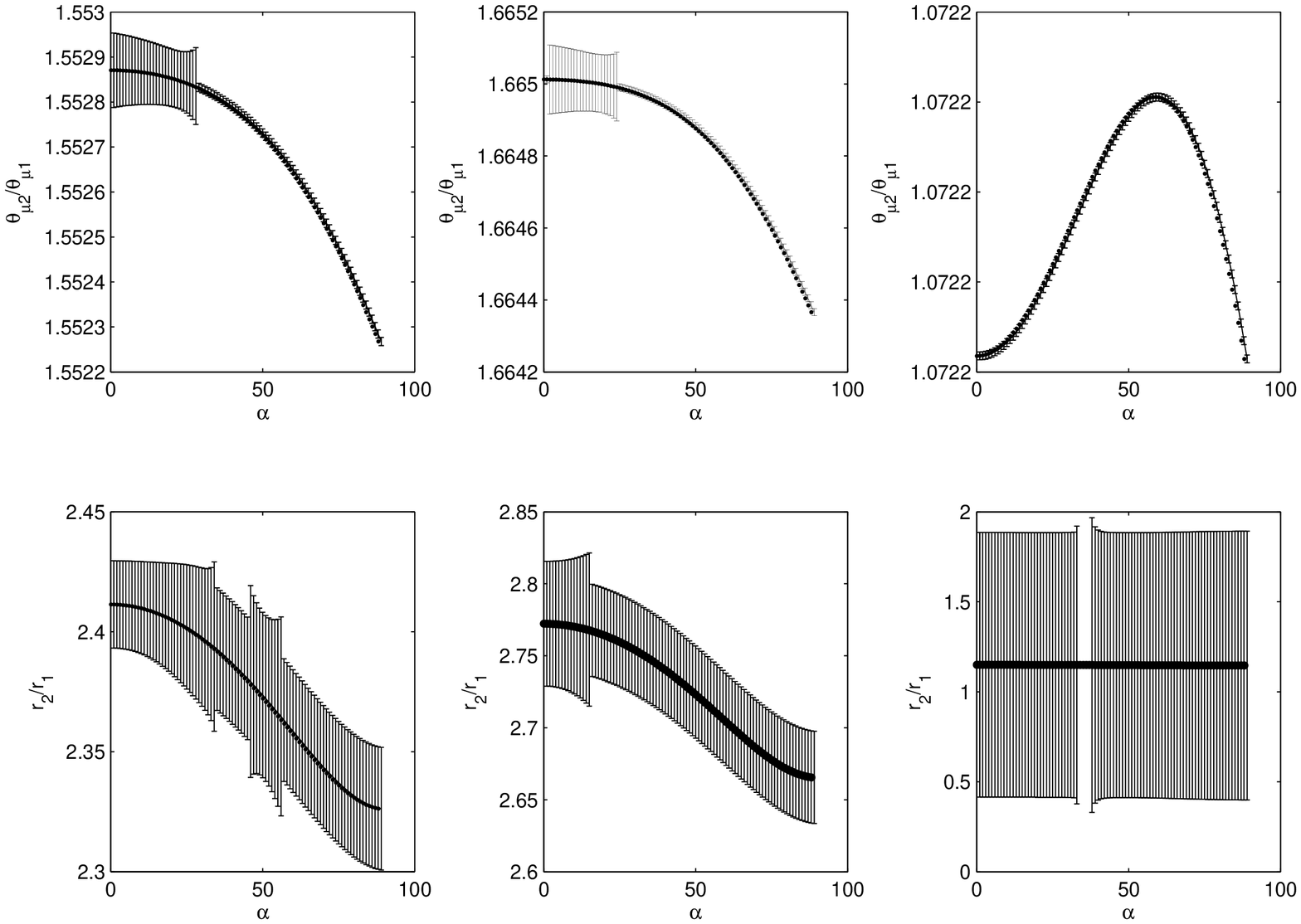}
\end{center}
\caption{$\theta_{\mu2}/\theta_{\mu1}$ and $r_2/r_1$ as the function
of $\alpha$ for PSR B1642-03 The left, middle, and right panels are
the ratios of 4.75GHz to 1.42GHz,  10.55GHz to 1.42GHz, and 10.55GHz
to 4.75GHz. The error bars indicate the error passing from the
errors of $\Delta\phi$ and $\kappa$. \label{fig:3}}
\end{figure}
The error comes from the $\Delta\phi$ (Kramer 1994) and
$\Delta\kappa$. (Here, we suppose $\Delta\kappa \approx 0.1\kappa$.)

From above two figures, we can see that the ratios are almost
constants when $\alpha$ change from $0^{\circ}$ to $90^{\circ}$.
This result can also be found in the Table 1, which shows the
variance $\Delta\eta$ to the mean $\bar{\eta}$. Here,
$\eta_1=\theta_{\mu2}/\theta_{\mu1}$, $\eta_2=r_2/r_1$,
$\Delta\eta=\eta_{\mathrm{max}}-\eta_{\mathrm{min}}$. It can be seen
that the variance is rather small comparable to the mean value. Thus
the ration $\theta_{\mu2}/\theta_{\mu1}$ and $r_2/r_1$ are
insensitive to $\alpha$ at all.
\begin{table}[]
 \begin{center}
  \begin{tabular}{c|ccc|ccc}
  \hline\noalign{\smallskip}
  \hline
  & & PSR B0950+08 & &  & PSR B1642-03 & \\
  \hline
 ratio &  $\eta_a$  & $\eta_b$   & $\eta_c$  &   $\eta'_a$  & $\eta'_b$   & $\eta'_c$ \\
  \hline
$\Delta \eta_{1}/\bar{\eta_{1}}$ & $5.5\times10^{-3}$  & $1.2\times 10^{-2}$ &  $1.8\times 10^{-2}$ & $3.9\times 10^{-4}$ & $3.9\times10^{-4}$ & $1.3\times10^{-5}$   \\
$\Delta \eta_{2}/\bar{\eta_{2}}$  & $1.5\times 10^{-2}$ & $1.6\times 10^{-2}$ & $2.3\times 10^{-2}$ & $3.8\times 10^{-4}$ & $3.9\times10^{-4}$ & $3.3\times10^{-3}$  \\
\hline
  \end{tabular}
 \end{center}
\caption{The ratio of the variance $\Delta\eta$ to $\bar{\eta}$ for
PSR B0950+08 and PSR B1642-03. The $\eta_a$, $\eta_b$, and $\eta_c$
are the ratios of 1420 MHz to 4750 MHz, 1420 MHz to 10.55 GHz, and
4750 MHz to 10.550 GHz of PSR B0950+08, respectively. The $\eta'_a$,
$\eta'_b$, and $\eta'_c$ are the ratios of 1430 MHz to 4750 MHz,
1430 MHz to 10.55 GHz, and 4750 MHz to 10.55 GHz of PSR B1642-03,
respectively.} \label{Tab:publ-works}
\end{table}
%\section

\section{The Parameter Space of ICS model}

The ICS model proposed by Qiao (1988) is model including the process
of a low frequency wave with angular frequency $\omega_0$ produced
in sparking and scattered by high frequency secondary particles with
Lorentz factor $\gamma$. Thus the observed angular frequency is
\begin{equation}
\omega'\simeq 2\gamma^2\omega_0(1-\beta_0\cos\theta_\mathrm{i}),
\label{Qiao}
\end{equation}
where $\beta_0=\nu/c\simeq 1$ as the secondary plasma moving with
relativistic velocity, $\theta_\mathrm{i}$ is the angle between the
direction of motion of low energy photon and high energy particle as
described by Qiao \& Lin(1998).

Applying observation and the calculations in Sect.2, we constrain
the parameter space of ICS model as follows. Firstly, we assume that
all the emission come from the last open field line and the Lorentz
factor $\gamma$ is a constant for a observed frequency \footnote{The
ICS model in Qiao (1988) does not assume a constant $\gamma$, here
we assume a constant $\gamma$ for simplicity.}. Secondly, because
one sparking process has time scale about $10^5-10^6$ s, the range
of $\omega_0$ is in $10^5-10^6$ s$^{-1}$. The secondary particles
have the energy with $\gamma\sim10^2-10^4$. Thirdly, according to
the assumption of Qiao \& Lin (1998) and Eq.\ref{Qiao}, we calculate
the theoretical ratios of beam width and emission height as function
of $\alpha$. Then we compared the simulated ratio from ICS model
with the ratio measured from observation by above method to
constrain the range of parameters $\omega_0$ and $\gamma$. The
results of PSR B0950+08 and PSR B1642-03 are shown in Fig.4 and
Fig.5, respectively.
\begin{figure}
\begin{center}
\includegraphics*[width=.8\textwidth]{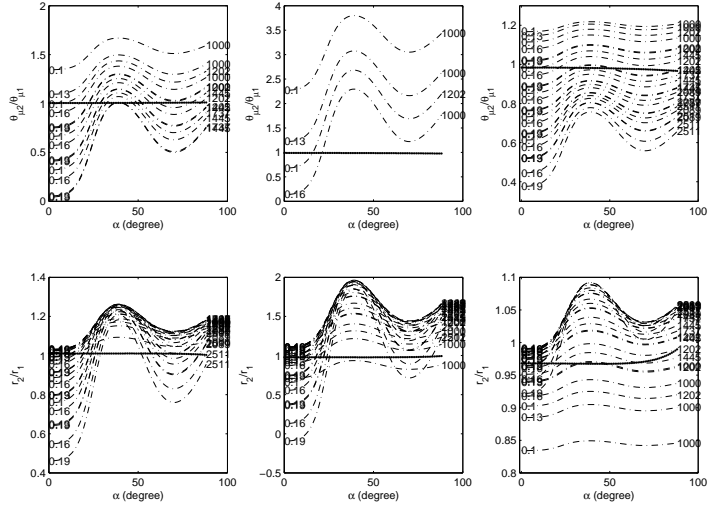}
\end{center}
\caption{Simulated and observed ratios $\theta_{\mu2}/\theta_{\mu1}$
and $r_2/r_1$ as the function of $\alpha$ for PSR B0950+08 The left,
middle, and right panels are the same ratios as in Fig.2. The black
dot line is the observed result from the pulse width. The dash dot
lines are the theory results from the ICS model. Each line has
different values of $\omega_0$ and $\gamma$, which are labeled in
the left and the right of the line, respectively. \label{fig:4}}
\end{figure}
\begin{figure}
\begin{center}
\includegraphics*[width=.8\textwidth]{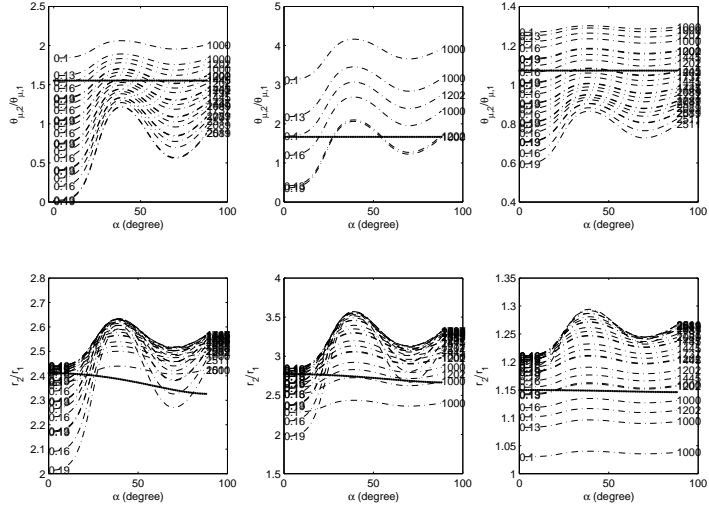}
\end{center}
\caption{Theory and observed ratios $\theta_{\mu2}/\theta_{\mu1}$
and $r_2/r_1$ as the function of $\alpha$ for PSR B1642-03 All the
denotations are the same as those of Fig.4. \label{fig:5}}
\end{figure}

From the figures, we can find: 1). The values of $\omega_0$ and
$\gamma$ can be fixed into a parameter space, $\omega_0 \in
[10^5,2\times10^5] \mathrm{s}^{-1}$ and $\gamma \in [10^3-2.5\times
10^{3}]$. 2). The constant $\gamma$ ICS model works better at high
frequencies, this may indicate that the Lorentz factor varies slower
at the higher location than at lower location.

\section {Conclusion and Discussion}

Using the observed pulse widths at different frequencies, we show
that the ratios of beam width and emission height between different
frequencies for PSR B0950+08 and PSR B1642-03 are insensitive to
inclination angle $\alpha$. A constant ratio indicates that $\alpha$
does not remarkably affect these ratios. These constants derived
from observation can constrain the parameter space of ICS model.
The comparison of observation and simulated results show that the
ICS model might work better in high frequencies than that in low
ones.

In the work, we only give a roughly parameter space (see Fig.6 for
details) but not give the exact values for the parameters. We hope
that this work can give hints to the future work to determine the
exact value of parameters about pulsar's radiation geometry and to
test more pulsar's radiation model.

\begin{figure}
\begin{center}
\includegraphics*[width=.8\textwidth]{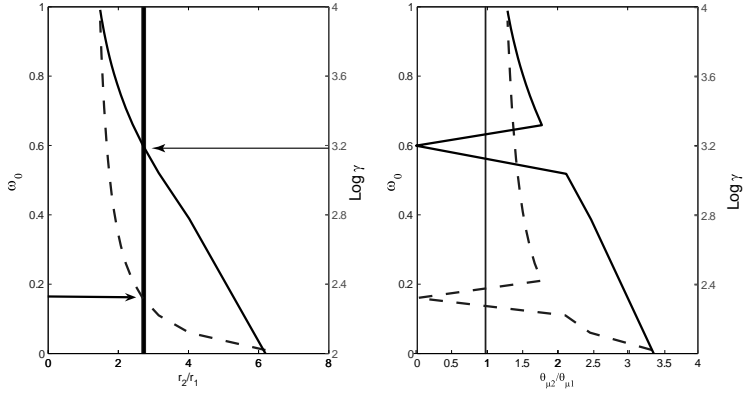}
\end{center}
\caption{$\omega_0$ and $\gamma$ as a function of $r_2/r_1$ for PSR
B1642-03 (left panel) and $\omega_0$ and $\gamma$ as a function of
$\theta_{\mu2}/\theta_{\mu1}$ for PSR B0950+08 (right panel). The
dash and the solid curves denote the relations of $\omega_0$ vs
$r_2/r_1$ and $\gamma$ vs $r_2/r_1$. Middle black bar is the
observed range. Two arrow marks give the exact values of $\omega_0$
and $\gamma$. Left panel gives the exact values for this pulsar:
$\omega_0=0.15$ and $\gamma=2.5\times10^{3}$, which is consistent
with the result shown in Fig.5. However, on the most of the
frequencies, $\omega_0$ and $\gamma$ as the function of $r_2/r_1$ or
of $\theta_{\mu2}/\theta_{\mu1}$ are not monotonic functions as
shown in right panel. Therefore, it is hard to give the exact value
but only the parameter space. \label{fig:6}}
\end{figure}

\end{document}